\documentclass[a4paper,prb,twocolumn,groupedaddress,showpacs]{revtex4}

\usepackage[utf8]{inputenc}
\usepackage{amssymb}
\usepackage{amsmath}
\usepackage{color}
\usepackage{graphics, graphicx}
\usepackage{bbold}
\usepackage{nicefrac}
\usepackage{mathcomp}
\usepackage{subfigure}
\usepackage[colorlinks=True, linkcolor=blue, citecolor=blue, urlcolor=blue]{hyperref}
\usepackage{color}
\usepackage{soul}

\newcommand{\ba}{BaFe$_2$As$_2$}
\newcommand{\otot}{O$\rightarrow$T}
\newcommand{\ttoct}{T$\rightarrow$cT}

\begin{document}

\author{Steffen Backes}
\affiliation{Institut f\"ur Theoretische Physik, Goethe-Universit\"at Frankfurt, Max-von-Laue-Str. 1, 60438 Frankfurt am Main, Germany}
\author{Harald O. Jeschke}
\affiliation{Institut f\"ur Theoretische Physik, Goethe-Universit\"at Frankfurt, Max-von-Laue-Str. 1, 60438 Frankfurt am Main, Germany}

\date{\today}

\title{Finite temperature and pressure molecular dynamics for BaFe$_2$As$_2$}

\begin{abstract}
  We study the temperature and pressure dependence of the structural
  and electronic properties of the iron pnictide superconductor
  {\ba}. We use density functional theory based Born-Oppenheimer
  molecular dynamics simulations to investigate the system at
  temperatures from $T=5$~K to 150~K and pressures from $P=0$~GPa to
  30~GPa. When increasing the pressure at low temperature, we find the
  two transitions from an orthorhombic to a tetragonal and to a collapsed
  tetragonal structure that are also observed in zero temperature
  structure relaxations and in experiment. However, these transitions
  are considerably smeared out at finite temperature, whereas the
  critical pressure for the first transition increases with
  temperature. We also analyze the electronic structure of {\ba} at
  finite temperature and work out differences between the time
  averaged band structure and Fermi surface at finite temperature
  compared to the known zero temperature results.  Our results should
  be helpful for resolving some open issues in experimental reports
  for {\ba} under high pressure.
\end{abstract}

\maketitle

\section{INTRODUCTION}
\label{sec:introduction}
The discovery of superconductivity in FeAs based compounds in 2008
with a critical temperature $T_c$ up to
$26$~K~\cite{KamiharaWatanabe2008} was the beginning of a large and
fast-evolving field dedicated to these new types of superconductors.
These materials like for example LiFeAs~\cite{TappTang2008,
  WangLiu2008}, SmFeAsO~\cite{Ren2008} and
{\ba}~\cite{RotterTegel2008Lett} share the structural motif of FeAs
layers in which superconductivity can emerge either by doping and/or
by external pressure with critical temperatures of up to $55$~K.  One
material of great interest is the compound {\ba} of the 122-type
family, which shows interesting behavior under external pressure and
temperature. At room temperature and ambient pressure, the
crystallo\-graphic structure of BaFe$_2$As$_2$ is that of the
body-centered tetragonal ThCr$_2$Si$_2$-type structure (space group
$I\,4/mmm$) with poor Pauli-paramagnetic metallic behavior and high
electrical resistivity~\cite{WangWu2009}. Upon lowering of the
temperature, the material undergoes a structural and magnetic
transition to the low-temperature orthorhombic $F\,mmm$
structure~\cite{RotterTegel2008} with a stripe-ordering of the Fe
magnetic moments~\cite{HuangQui2008_2}.  External
hydrostatic or uni-axial pressure is capable of suppressing the
magnetically ordered orthorhombic phase and induces the tetragonal
paramagnetic phase with a region of
superconductivity~\cite{AlirezaKo2008,ColombierBudko2009,PaglioneGreene2010}
at low temperatures. At higher pressures, the collapsed tetragonal phase
emerges~\cite{UhoyaStemshorn2010,MittalMishra2011}, where experiments
as well as recent theoretical investigations~\cite{Tomic2012} find
different values for the critical pressure, which seems to be
sensitive to temperature. At $300$ K, values of $16.7$~GPa under
nonhydrostatic pressure conditions~\cite{UhoyaStemshorn2010} and 
$27$~GPa~\cite{MittalMishra2011} under hydrostatic pressure conditions
are reported, whereas at $33$ K a critical pressure of $29$ GPa was
found~\cite{MittalMishra2011}. Also hysteresis effects of the
temperature or a coexistence of the orthorhombic and tetragonal phase
might be present for a wide range of
pressures~\cite{MittalMishra2011}.  These experimental results show
that the critical pressures and their possible temperature dependence
are not yet fully understood.

In recent years, there have been quite a few theoretical studies of
structural effects in BaFe$_2$As$_2$. The phase transitions at zero
temperature were studied with Car Parrinello molecular dynamics with
friction~\cite{Zhang2009}, highlighting the importance of Fermi
surface nesting for the nature of the phase transitions. A combination
of DFT total energies and thermodynamical quantities estimated from
experiment was used to rationalize some finite temperature transition
points in the phase diagram~\cite{JiLu2011}. Constant volume DFT
relaxations have been used to study the zero temperature
pressure-induced structural transitions and equation of state both for
hydrostatic~\cite{Colonna2011a} and nonhydrostatic~\cite{Colonna2011b}
pressure conditions.
The phase transitions in {\ba} have also been studied using the fast
inertial relaxation engine~\cite{Tomic2012}, which allows an
unconstrained equilibrium structure search for arbitrary stress
tensors. This method has been used to work out effects of uniaxial
pressure along {\bf c} direction~\cite{Tomic2012} and also in the $ab$
plane~\cite{Tomic2013}.  Since most studies so far used either
experimentally measured or zero-temperature optimized structures, the
complex temperature and pressure phase diagram of {\ba} calls for
microscopic first principles calculations at finite temperature.

Therefore, in this study we extend the theoretical toolbox for the
study of {\ba} to include finite temperatures.  We employ \textit{ab
  initio} density functional theory (DFT) methods combined with finite
temperature and pressure molecular dynamics.  We specifically
investigate the temperature behavior of the structural transitions and
electronic properties.  To our knowledge, no theoretical
investigations of the finite temperature regime with external pressure
of {\ba} have been carried out before.  Therefore, our results can
provide a useful complement to existing $T=0$~K theoretical
calculations and can help to clarify the contradicting observations
that are reported for the transition pressures in experiments.

\section{METHODS}
\label{sec:methods}

\begin{figure*}[t]
\begin{center}
\includegraphics[width=0.95\textwidth]{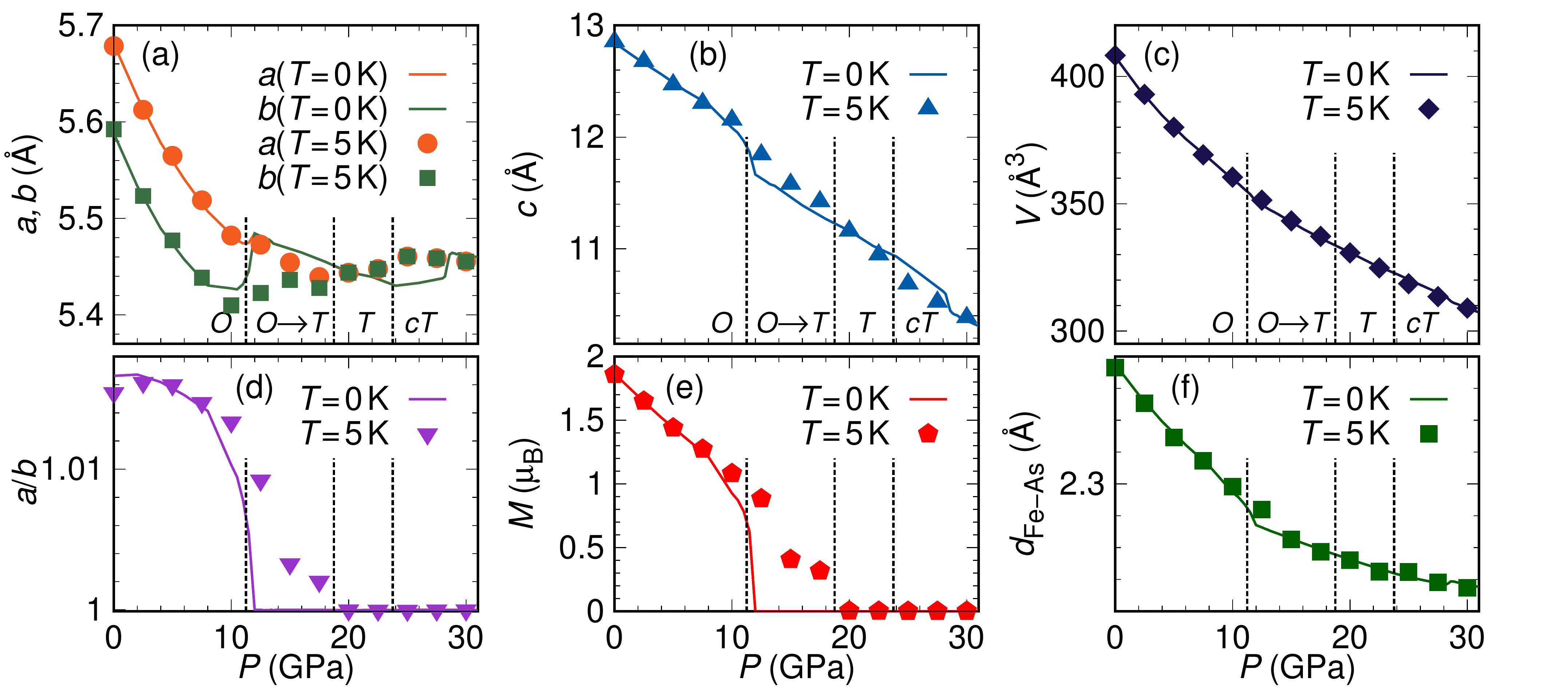}
\end{center}
\caption{(Color online) Structural parameters for {\ba} at $T=5$ K as a function of
  pressure (symbols).  (a) Time averaged $a$ and $b$ lattice
  parameters, where $a$ is in the direction along the AFM ordering of
  the magnetic moments. (b) Time averaged $c$ lattice parameter. (c)
  Volume of the unit cell comprising $8$ Fe atoms.  (d) $a/b$ ratio,
  (e) absolute value of the (time and unit cell averaged) Fe magnetic
  moment and (f) the Fe-As distance, averaged over time and all
  nearest neighbor Fe-As bonds.  The solid line is a $T=0$ K
  optimized structure from Ref.~\onlinecite{Tomic2012}. }
\label{res:lattice_parameters}
\end{figure*}

We performed the molecular dynamics calculations using the atomic
simulation environment (ASE)~\cite{ASE} interface combined with
the DFT Vienna ab-initio simulation package (VASP)~\cite{Vasp},
version 5.2.11, with the projector augmented wave
basis~\cite{pawPotentials} in the generalized gradient approximation
(GGA) by Perdew, Burke and Ernzerhof~\cite{GGAPBE}. To control the
temperature and pressure, the Berendsen dynamics~\cite{Berendsen}
provided by ASE were employed, using characteristic time-constants of
$\tau_T=50$~fs, $\tau_P=250$~fs for the thermostat and barostat,
respectively, and a time step of $1$~fs for the integration of the
equations of motion. The energy cut-off in VASP was set to $400$~eV
and a Monkhorst-Pack uniform grid of $(4\times4\times4)$ k-points was
used for the integration of the Brillouin zone. To account for the
magnetic stripe-order, we simulate a $\sqrt{2}\times\sqrt{2}\times 1$
supercell of the conventional orthorhombic unit cell of {\ba}
containing 8~Fe atoms. Calculations were performed on a
pressure-temperature grid with 13 different values of $P$ going from
$0$~GPa to $30$~GPa in $2.5$~GPa steps and for temperatures of $5$~K,
$50$~K, $100$~K and $150$~K.  The values of observables were obtained
by averaging over the last $400$~fs of the trajectory after the
configuration was sufficiently equilibrated.

\section{RESULTS}
\label{sec:results}
\subsection{Effect on transitions}
\label{subsec:res:transition}

We now present our results for the lattice parameters, magnetic
moments and volume at temperatures between $5$~K and $150$~K compared
to a zero-temperature relaxation~\cite{Tomic2012}. In 
Fig.~\ref{res:lattice_parameters} we show the pressure dependence of the
three lattice parameters $a$, $b$ and $c$ at $T=5$~K, which is very similar
to the $T=0$ K case. We observe a transition from the orthorhombic to
tetragonal structure at about $12.5$ GPa and a second transition from
the tetragonal to a collapsed tetragonal structure which is around
$25$ GPa.  For low values of pressure, \textit{i.e.} \mbox{$0-5$ GPa},
our results are almost identical to the $T=0$ K calculation, with our
obtained values for the $a$ and $b$ axis being slightly longer in
comparison. The reason for that is clearly given by finite
temperature which leads to internal pressure due to the temperature
fluctuations of the atomic position and thus increasing the volume of
the unit cell compared to the $T=0$ K case.

The position of the transition to the tetragonal phase is
visible as a sudden increase in the lattice parameter $b$ at around
$12.5$ GPa. The position of the transition is in agreement with the
$T=0$ K result but the resulting configuration is not a perfect
tetragonal unit cell, with the orthorhombic distortion significantly
reduced but still non-zero.  The $\frac{a}{b}$-ratio drops from
$1.013$ at $10$ GPa to $1.009$ at $12.5$ GPa and retains a value
larger than one up to $17.5$ GPa.


Even though all the lattice parameters noticeably differ from the
$T=0$ K results in some cases, our obtained volume of the unit cell is
still very similar (see Fig.~\ref{res:lattice_parameters} c)).

For increasing pressure we find the average distance of the iron and
arsenic atoms in the Fe-As tetrahedron layers to be reduced and the
orthorhombic transition is visible as a kink in the Fe-As distance
plot. Our results closely match the zero temperature calculation and
correctly indicate the orthorhombic to tetragonal transition around
$12.5$ GPa (see Fig.~\ref{res:lattice_parameters} (f)).

The tetragonal to collapsed tetragonal transition at zero temperature
is marked by a sudden increase of the $a$ and $b$ and a decrease in
the $c$ lattice constant at $27.5-30$ GPa. Our results show a
smoothed-out transition with less drastic changes in the lattice
parameters at $P=22.5-25$ GPa. The $a$ and $b$ lattice parameters
increase by $0.23\%$ from $22.5$ GPa to $25$ GPa, whereas the increase
for $T=0$ K is $0.66\%$ from $27$ GPa to $30$ GPa. The $c$ lattice
parameter shows a similar behavior, where in the zero temperature case
a distinct decrease at $27.5-30$ GPa is visible, but our results show
a smooth behavior of $c$ that looks almost linear with no sharp drop.

The most notable difference between finite temperature and previous
zero temperature results is the non-tetragonal but intermediate
orthorhombic-tetragonal structure of our system for values of pressure
where the $T=0$ K structure is already completely tetragonal.  Thus,
our calculations obtain a smooth transition rather than sudden change
of the lattice parameters for the {\otot} transition and also for the
{\ttoct} transition.  This is to be expected since due to the
temperature fluctuations the system is allowed to oscillate between
two competing structural configurations and magnetic orderings close
to the critical pressure at $T=0$ K. Our result suggests that this
phase, which is intermediate between orthorhombic and tetragonal phases,
is the best approximation we can get with our 20 atom supercell to a
mixed phase. As a result, the {\otot} transition is smoothed out to
higher pressures, whereas the {\ttoct} transition  experiences strong 
smearing effects. This behavior is strongly enhanced for higher temperature,
where the {\otot} transition is shifted to even higher pressures of 
about $15$~GPa and the {\ttoct} transition becomes almost indiscernible.
This gives interesting effects in the relative change of the lattice 
parameters with temperature, which we will discuss below.

\begin{figure}[t]
\begin{center}
\includegraphics[width=0.45\textwidth]{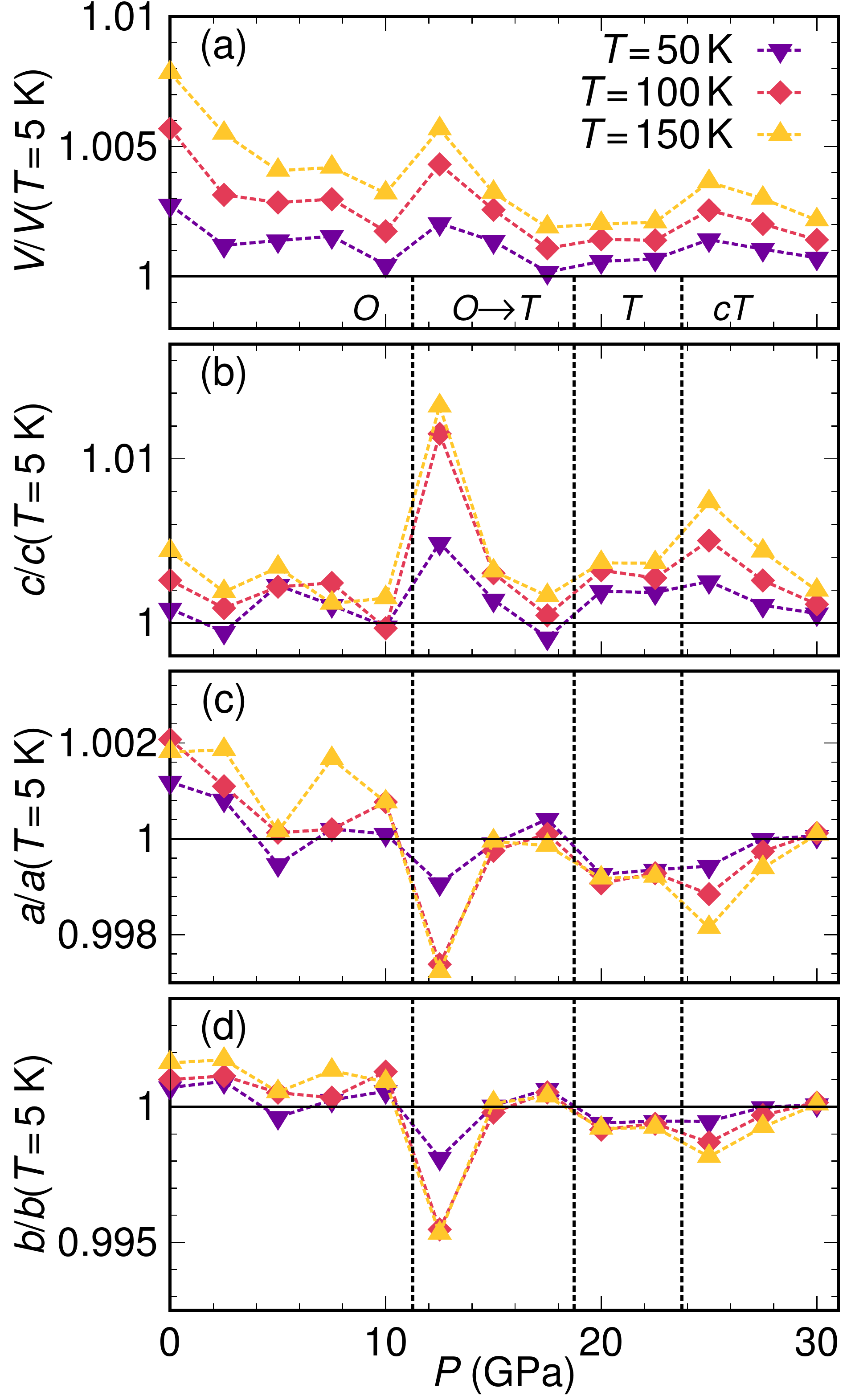}
\end{center}
\caption{(Color online) The relative increase of the volume $V$ of the unit cell and
  the lattice parameters $a$,$b$ and $c$ compared to $T=5$ K at
  different temperatures as a function of external pressure. The
  dashed vertical lines indicate the critical pressure that we
  identified for the corresponding structural transitions at
  $T=5$~K. Note that in particular the first phase transition shifts
  to higher pressures as the temperature is increased.}
\label{res:thermal_exp}
\end{figure}

The increase of the volume of the unit cell and the lattice parameters
$a$, $b$, $c$ with temperature can be seen in
Fig.~\ref{res:thermal_exp}.  For $50$ K the volume increases slightly
at all pressures compared to $T=5$ K, with different behavior at the
critical pressures. Up to the first transition, the increase of the
volume with temperature decreases as expected due to higher pressure and 
then shows a peak right at $12.5$ GPa, which is caused by the temperature
induced shift of the {\otot} transition, \textit{i.e.} 
the sudden decrease in volume, to higher pressures.  In all of the regions we
identified, except in the tetragonal phase the relative volume increase falls 
off almost monotonically until the transition to the next phase, where it 
shows a small jump to higher values, after which it decreases again. Only 
in the tetragonal phase the increase with temperature seems to be pressure independent.
Comparatively, the jump of the relative volume increase is largest at the high pressure
end of the orthorhombic phase.  This behavior is mirrored in the
temperature dependence of the $c$ lattice parameter at the critical
pressures. We observe that at a pressure of $P=12.5$~GPa the relative
change of $c$ when increasing the temperature from $T=5$~K to $50$~K
and from $T=50$~K to $100$~K compared to $T=100$~K to $150$~K is considerably larger.
This can be explained by the temperature-induced transition from the tetragonal
phase to the orthorhombic phase. This is also the reason for the
relative decrease of the $a$ and $b$ lattice parameters at $12.5$ GPa.  Thus,
we obtain a positive slope of the orthorhombic to tetragonal phase
transition in the temperature-pressure phase diagram of {\ba}.

\begin{figure}[t]
\begin{center}
\includegraphics[width=0.45\textwidth]{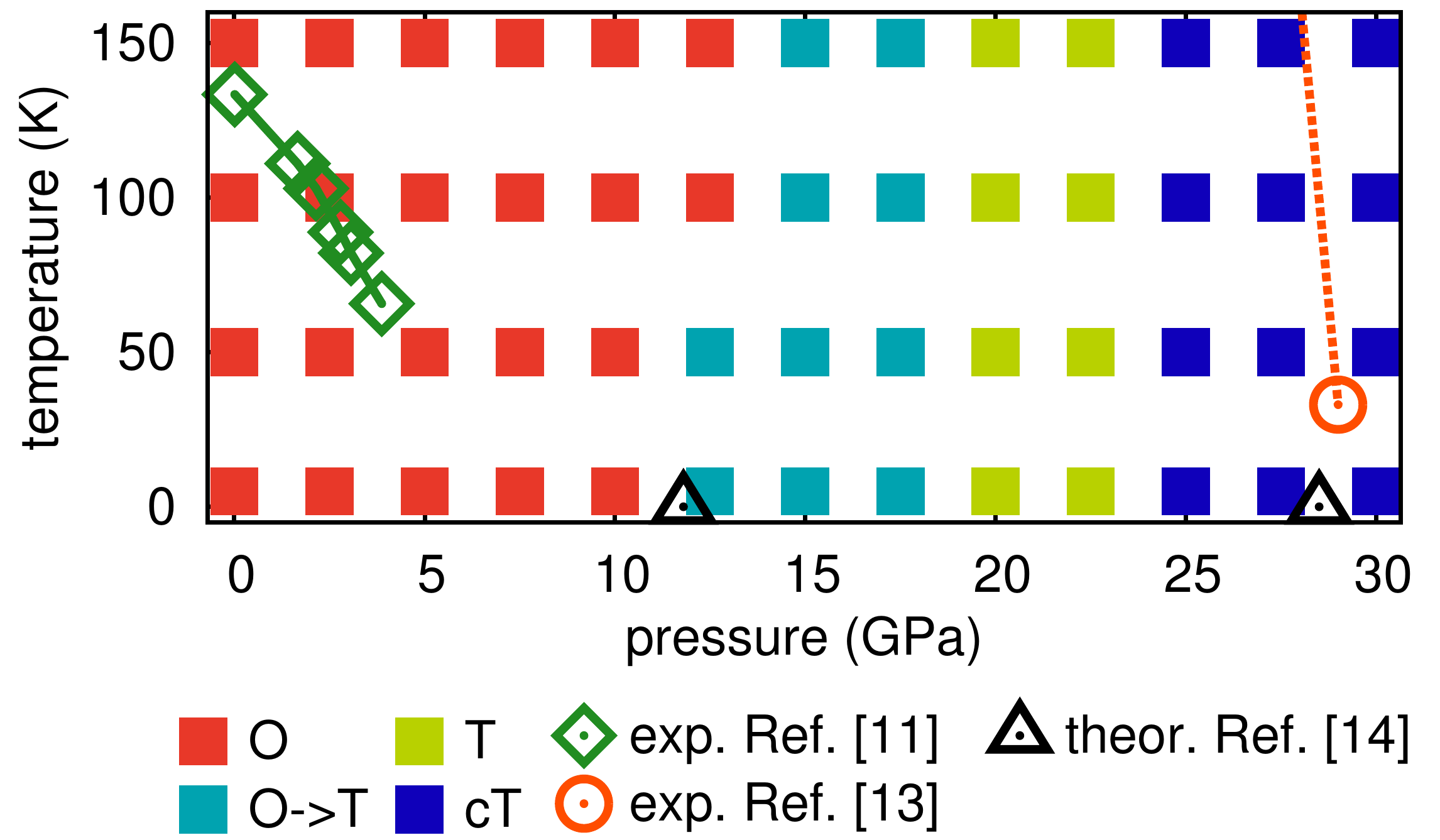}
\end{center}
\caption{(Color online) The temperature-pressure phase diagram of BaFe$_2$As$_2$,
  where we distinguish between four different phases (see detailed
  explanation in the text). Experimental data points by Colombier
  \textit{et al.}~\cite{ColombierBudko2009} (green line) represent the
  {\otot} transition and the data by Mittal \textit{et
    al.}~\cite{MittalMishra2011} (orange symbols) represents the
  {\ttoct} transition.  The two black data points obtained by a $T=0$
  K relaxation by Tomi\'{c} \textit{et al.}~\cite{Tomic2012}
  also indicate the two critical pressures.}
\label{res:phasediagram}
\end{figure}

Finally, we set up a temperature pressure phase diagram by specifying
the state of the system for each $(T,P)$ combination we calculated,
\textit{i.e.} by taking into account if the structure is orthorhombic
or tetragonal, the magnetic moments, \textit{etc.}  To determine the
transition pressures used in the diagram we investigated the
pressure-derivative of the $a,b$ and $c$ lattice parameters and the
volume. 
In the phase diagram (see
Fig.~\ref{res:phasediagram}) we distinguished four different regions
of structural configurations: The first one is the phase of the purely
orthorhombically distorted unit cell, in which all the lattice
parameters still continue to decrease monotonically with higher
pressure. As we have seen, this phase is present for pressures up to
$P=10$~GPa for the temperatures $T=5$~K and 50~K and extends to
$P=12.5$~K for the two higher temperatures.  The second region is an
intermediate orthorhombic-tetragonal state, in which the shorter axis
$b$ of the unit cell is no longer decreasing but increases with higher
pressures and has not reached the same length as the longer axis
$a$. This region is quite prominent in our results since the
transition is always smoothed out considerably compared to the $T=0$ K
result~\cite{Tomic2012}.  The next phase is the one of a pure
tetragonal unit cell with equal lattice parameters $a=b$.  After that,
at high pressures the collapsed tetragonal phase is dominating, which
is defined by the sudden increase in the $a$ and $b$ and the decrease
in the $c$-axis.
By comparing the temperature dependence of the lattice parameters at
fixed values of the pressure we suspect the {\ttoct} transition to be
pushed to lower pressures at higher temperature, which was found by
Mittal \textit{et al.}~\cite{MittalMishra2011}, but due to strong
smearing effects we could not determine a slope in the critical pressure.
However, we find the transition to the collapsed tetragonal phase to
occur at a pressure being $5-7.5$~GPa lower compared to experiments.

\subsection{Band Structure}
\label{subsec:res:bandstructure}
In order to investigate the effect of the finite temperature
fluctuations on the electronic properties of BaFe$_2$As$_2$, we
performed band structure calculations using the full-potential
local-orbital minimum-basis code (FPLO)~\cite{FPLO}.  For the
calculation a $6\times 6 \times 6$ {\bf k} mesh was employed and the
generalized gradient approximation (GGA) method by Perdew, Burke and
Ernzerhof~\cite{GGAPBE} was used for the exchange- and correlation
potential.  The tetrahedron-integration method was used for the
integration of the Brillouin zone and the density was
converged with an accuracy of $10^{-6}$.

\begin{figure*}[t]
\includegraphics[width=0.7\textwidth]{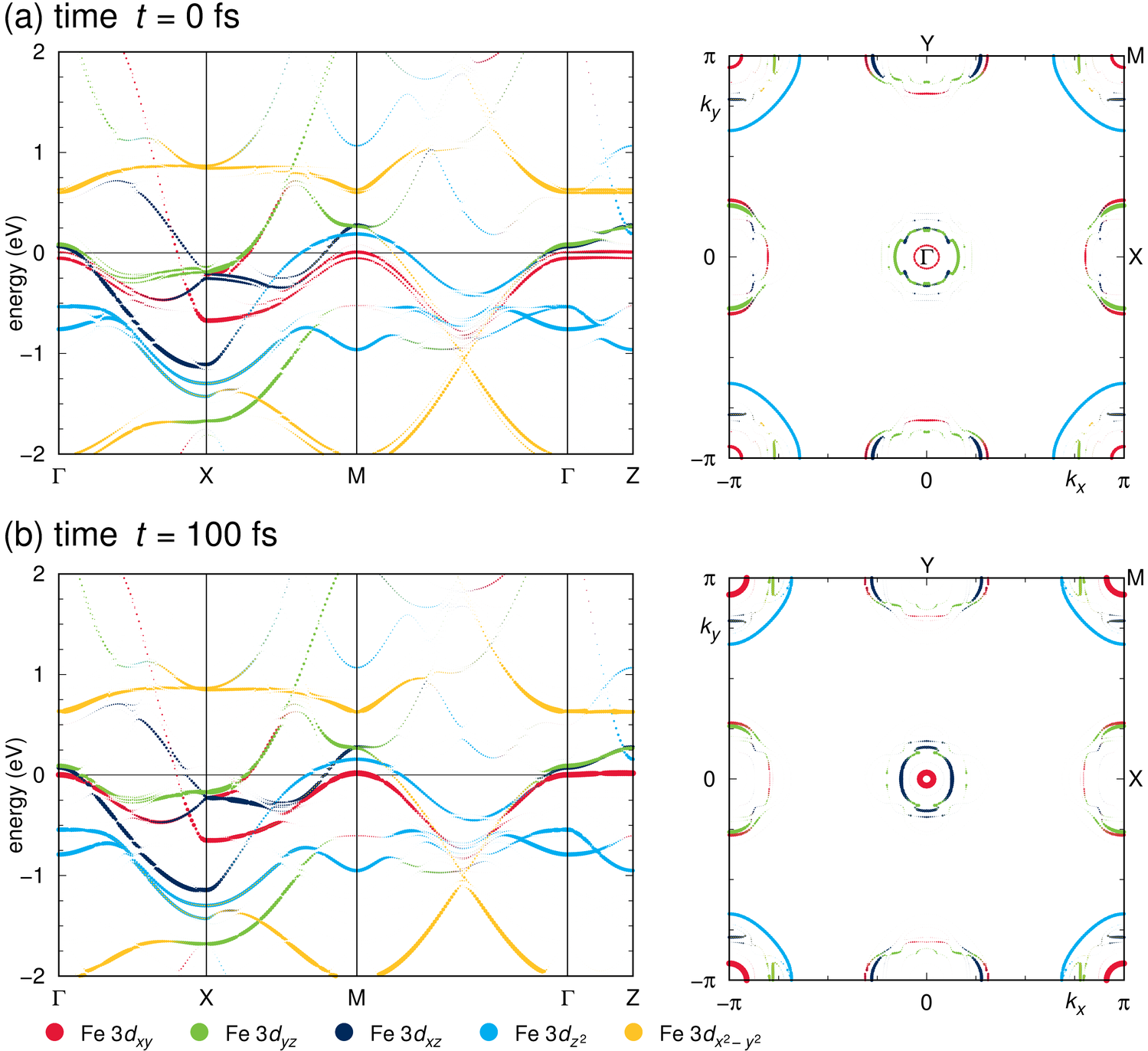}
\caption{(Color online) Band structure of {\ba} and $k_z=0$ Fermi surface cut at 
$T=100$~K and different times within one MD trajectory. 
}
\label{res:bands_comp}
\end{figure*}

As the first configuration we investigated a temperature of $T=100$~K
and pressure of $P=0$~GPa, where an orthorhombic unit cell with non-zero
magnetic moments is present. At this temperature the volume is
periodically fluctuating around a mean value of $410.747$
$\mathring{\mathrm{A}}^3$ $\pm 0.07\%$.  The duration for a complete
oscillation cycle of the volume is about $210$~fs. We took snapshots
of the crystal structure in $10$~fs intervals, starting from the
maximal expanded volume down to the minimum value of the volume, so in
total we covered one half of the oscillation cycle of about $100$~fs.
For the band structure we use the usual path in {\bf k}-space $\Gamma
= (0,0,0)$, $\mathrm{X} = (\nicefrac{1}{2},0,0)$, $\mathrm{M} =
(\nicefrac{1}{2},\nicefrac{1}{2},0)$, $\Gamma$, $\mathrm{Z} =
(0,0,\nicefrac{1}{2})$, given in units of the reciprocal lattice
vectors. We chose this path since it corresponds to the higher symmetry
of the Fe lattice rather than to the $F\,mmm$ space group and it
can be easily compared to the band structure of the more simple
tetragonal LaOFeAs~\cite{Mazin2008}.

\begin{figure*}[t]
\includegraphics[width=0.7\textwidth]{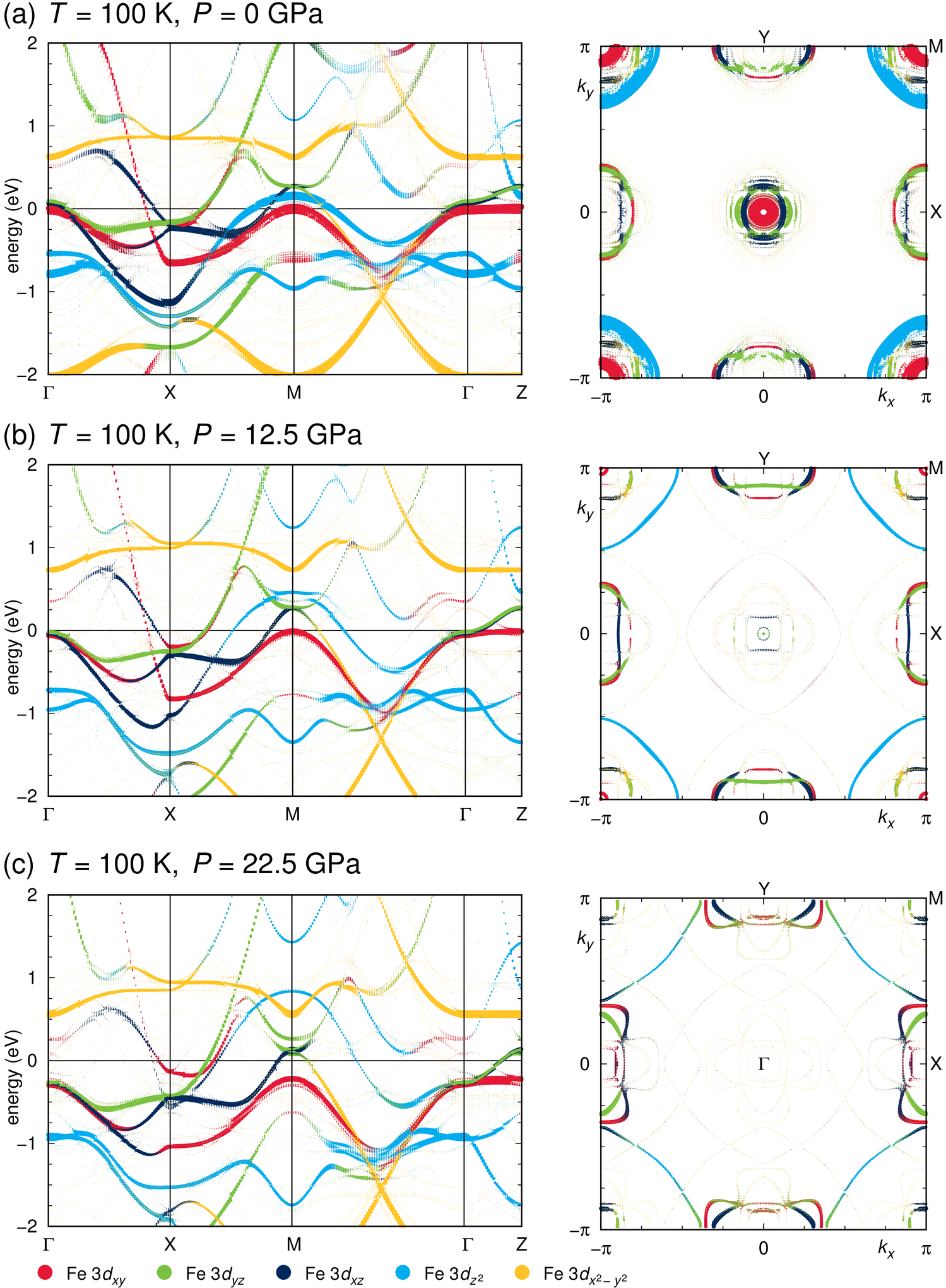}
\caption{(Color online) Time averaged band structures and $k_z=0$ Fermi surface cuts
  at a temperature $T=100$~K for three different pressures.  (a)
  $P=0$~GPa corresponds to the orthorhombic phase, (b) $P=12.5$~GPa is
  in the intermediate region between orthorhombic and tetragonal
  phases, and (c) $P=22.5$~GPa corresponds to the tetragonal
  phase. Averaging over several electronic structures along an MD
  trajectory leads to significant broadening of bands and Fermi
  surface contours, especially at low pressure.}
\label{res:overlay_bs}
\end{figure*}

We compared our results to a zero-temperature {\ba} structure at
$P=0$~GPa from ~\cite{Tomic2012} to investigate how the finite
temperature affects the bands near the Fermi level.  Since the most
important contribution to the DOS at the Fermi level in this materials
is usually given by the Fe bands, which we also confirmed to be the
case here, we will now discuss the effect of the finite temperature on
the Fe bands exclusively.  For a meaningful comparison one has to mind
that in the finite temperature case the positions of the atoms in the
unit cell are fluctuating and thus the eight Fe atoms are no longer
equivalent and the crystal structure is no longer the $F\,mmm$ space
group but rather triclinic $P\,1$. Due to the higher number of
inequivalent Fe atoms we also get additional Fe bands that are folded
back into the Brillouin zone. By using the band unfolding method that
was prosed in Ref.~\cite{Ku2010} and implemented in
FPLO~\cite{vanHeumen2011}, we can unfold the band structure and map
the equivalent bands onto a single one.

\begin{figure*}[t]
\includegraphics[width=0.95\textwidth]{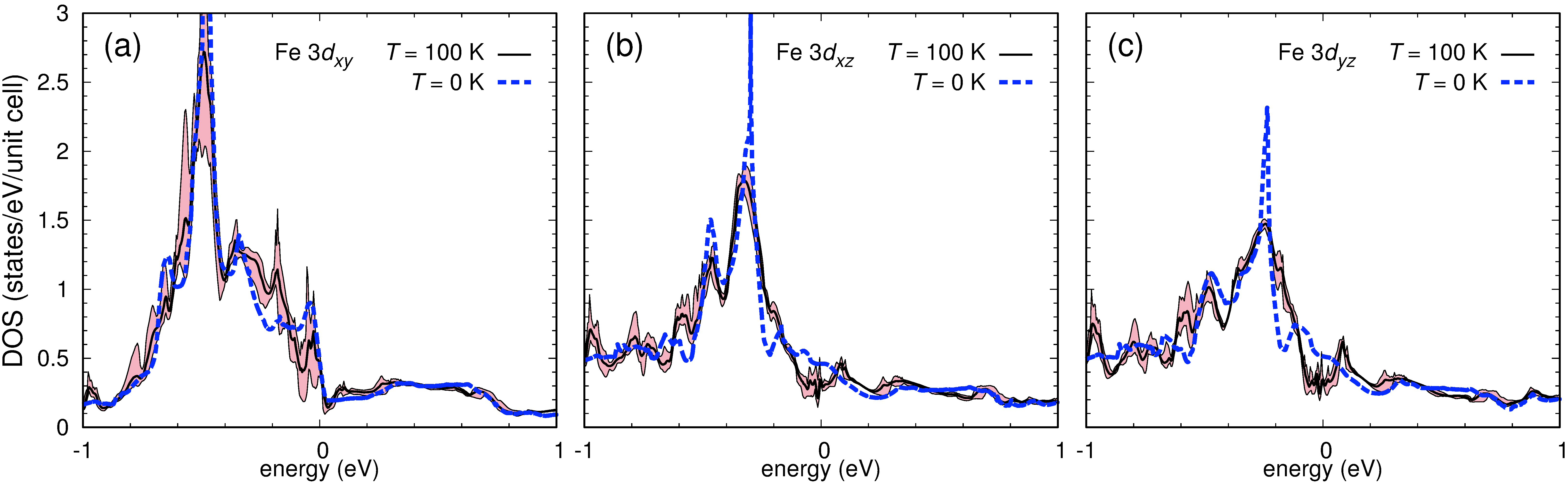}
\caption{(Color online) Partial densities of states, compared at $T=0$~K (dashed
  line)~\cite{Tomic2012} and $T=100$~K (solid line) for the three Fe
  $t_{2g}$ orbitals $3d_{xy}$, $3d_{xz}$ and $3d_{yz}$. The solid line
  is the fluctuation-averaged DOS and the colored area indicates the
  extent of the fluctuations due to temperature (see detailed
  explanation in the text).}
\label{res:overlay_dos}
\end{figure*}

In Fig.~\ref{res:bands_comp} we show the Fe orbital weighted
band structure and the $k_z=0$ Fermi surface cut of {\ba} at $T=100$~K
and ambient pressure.  The figure shows instantaneous electronic
structures for two structural configurations during a complete volume
oscillation cycle of the unit cell, separated by a time of $t=100$~fs.
The first configuration (Fig.~\ref{res:bands_comp}~(a)) corresponds to
the maximal volume and the second one (Fig.~\ref{res:bands_comp}~(b))
to the minimal volume during the oscillation cycle.  Close to the
Fermi level $E_F$ the most important bands are the Fe $3d_{xy}$,
$3d_{xz}$ and $3d_{yz}$ states. The unfolding procedure occasionally
leads to more crossings of the Fermi surface than at $T=0$ because in
their thermal motion the Fe sites are only approximately equivalent.
Fe $3d_{xy}$, $3d_{xz}$ and $3d_{yz}$ bands form hole pockets at the
$\Gamma$ point and in addition $3d_{z^2}$ character is present at the
$\mathrm{M}$ point. Electron pockets at the X (and Y) points are of
$3d_{xy}$, $3d_{xz}$ and $3d_{yz}$ character.  Compared to $T=0$~K
calculations, the effect of the finite temperature is clearly present
in the shallow inner hole pocket of $3d_{xy}$ character at the
$\Gamma$ point, which clearly shows size fluctuations during the
trajectory.
Shrinking of this hole pocket can be compared to the effect of
 hydrostatic and uniaxial
pressure~\cite{Tomic2012}, where the Fe $3d_{xy}$ band is
pushed below the Fermi level at the magnetic {\otot} transition.  
The structural variations along the MD trajectory at $T=100$~K also
show interesting effects on the outer hole cylinders at the $\Gamma$
point. While they appear elliptical along $k_y$ at $t=0$~fs, they
appear elliptical along $k_x$ at $t=100$~fs. This change of shape goes
hand in hand with changing importance of $3d_{yz}$ and $3d_{xz}$
characters: more $3d_{yz}$ at $t=0$~fs, more $3d_{xz}$ at
$t=100$~fs. This feature resembles the electronic structure changes
upon uniaxial compressive and tensile stress in the $ab$ plane of
{\ba} (compare Ref.~\onlinecite{Tomic2013}).  Other bands seem to be less
affected by the volume oscillations.

We now proceed from the consideration of transient shapes of the Fermi
surface to its thermal average. Transient features of the electronic
structure are interesting theoretically and can in principle, at least
following the excitation of coherent phonons, be observed in
experiments with very high temporal
resolution~\cite{Rettig2012,Avigo2013}. Since usual angle resolved
photo emission observes time averaged electronic structures, a comparison 
to the averaged band structures like the ones we show in 
Fig.~\ref{res:overlay_bs} might be more appropriate. Note, however, that
while our electronic structures include the effect of thermally
fluctuating crystal structures, they do not include the thermal
broadening of the spectral function which can be described by finite
temperature Greens functions.
Fig.~\ref{res:overlay_bs} shows the time averaged
band structure at $T=100$~K for the pressures $P=0$~GPa, $12.5$~GPa and
$22.5$~GPa, where the first one is deep in the orthorhombic phase, the
second is in the smooth transition region between orthorhombic and
tetragonal phases and the latter is in the tetragonal phase.

Our results clearly show that the Fe $3d_{z^2}$ band at the $M$ point
and the Fe $3d_{xy}$ band at the $\Gamma$ and $M$ points experience
noticeable changes due to thermal structure fluctuations. In the plot
the broadening of the bands is a lot larger than the line width used
to draw them, which has implications for interpreting and comparing
theoretical iron pnictide band structures and experimentally measured
spectral functions. Bands that are below the Fermi level in
theoretical zero-temperature calculations can become partially
occupied even at temperatures way below room temperature like
$T=100$~K just because of the structural fluctuations caused by
temperature.  This could have implications for the interpretation of
angle-resolved photo emission spectroscopy (ARPES) results as it leads
to an additional broadening in spectral functions, besides the usual
thermal broadening.

As the pressure is increased to $12.5$ GPa, which in our case is in
the orthorhombic-tetragonal smoothed transition region, we notice a
significant shift of bands away from the Fermi level as already
reported previously~\cite{Zhang2009,Tomic2012}. The Fe $3d_{xy}$ band
becomes fully occupied at the $\Gamma$ point and fluctuates around
$E_F$ at the $M$ point, whereas the Fe $3d_{z^2}$ band is shifted to
higher energies.  The finite Fe $3d_{xy}$ weight at the Fermi level at the
$M$ point is in contrast to the previous $T=0$ K
calculation~\cite{Tomic2012} and corresponds to the smooth {\otot}
transition.  Compared to the $P=0$~GPa result, the temperature induced
fluctuations are greatly reduced and only little smearing
of the Fermi surface is observed.

In the tetragonal phase at $22.5$ GPa the Fe bands are pushed further
away from the Fermi level, with no crossings at the $\Gamma$ point and
only small Fe $3d_{yz}$ and $3d_{xz}$ hole pockets at the $M$ point.
Thermal smearing of the Fermi surface is further reduced.

The effect of the temperature fluctuations in the Fe bands is also
visible in the density of states (DOS).  In Fig.~\ref{res:overlay_dos}
we show the Fe $3d$ density of states for the three most important Fe
$3d_{xy}$, $3d_{yz}$ and $3d_{xz}$ bands.  Compared to the zero
temperature case we find our previous results for the $T=100$ K
calculations to be confirmed also in the DOS.  The change of the DOS
for energies above $E_F$ is rather subtle while below we see a general
shift of weight to lower energies or an increase of the DOS compared
to $T=0$ K. Right at the Fermi level and slightly below the change is
strongest, where the DOS is significantly reduced. This lowering of
the density of states at the Fermi level $N(E_F)$ is due to a lowering
of the symmetry of the system at finite temperature which reduces the
energetically unfavourable large $N(E_F)$ (Stoner instability).  The
thermal fluctuations of the DOS are clearly visible with average
fluctuations of $\sim 15-20\%$ around the mean value below the Fermi
level and $\sim 5-10\%$ above.  Right at $E_F$ the fluctuations seem
to reach a local maximum and can approach large peak values of about
$\sim 40\%$, like in the Fe $3d_{xy}$ orbital at $-0.05$ eV.  For
higher pressures we find the fluctuations to be reduced as expected
and do not show them here.

\section{CONCLUSION}
\label{sec:conclusion}
We used density functional theory based Born-Oppenheimer molecular
dynamics to study the behavior of {\ba} under external pressure and
finite temperature.  For low temperatures, our results are in
agreement with existing experimental as well as other theoretical
results, correctly indicating the transition from the orthorhombic
magnetic structure to a tetragonal paramagnetic phase around $12.5$
GPa and the tetragonal to a collapsed tetragonal phase around $\sim
25$ GPa.  In general, the transitions were found to be considerably
smoothed out due to finite temperature and the critical pressure for
the {\otot} transition was shifted to higher pressures, whereas the
{\ttoct} transition becomes almost indiscernible. These effects became
more enhanced with higher temperature and we found a positive slope of the
{\otot} transition line in the temperature-pressure phase
diagram. Thus, our results confirm that temperature is an
important factor for the structural transitions in {\ba}.

We also investigated the electronic structure of {\ba} and the effect
of structural fluctuations at finite temperature on the band structure
and density of states. At $T=100$~K and $P=0$~GPa we find a
fluctuating size of the hole pockets at the $\Gamma$ and $\mathrm{M}$
symmetry points.  The Fe $3d_{xy}$ band shows strong oscillations at
$E_F$ with significant variations in the DOS, and Fe $3d_{xz}$ and
$3d_{yz}$ weights are periodically oscillating at the $\Gamma$ point
due to thermal structural fluctuations.  At ambient pressure, the
fluctuation in the band energies caused by temperature are significant
and lead to a thermal broadening of the the Fermi surface.  Therefore,
compared to zero temperature calculations which miss the dynamical
fluctuations and overestimate the DOS at the Fermi level, our results
indicate that thermal fluctuations cause non-negligible effects in the
electronic structure even at temperatures well below room temperature,
which could be important for the interpretation of experiments like
angle-resolved photo emission spectroscopy (ARPES).

\acknowledgments 

The authors would like to thank Roser Valent{\'\i} for useful
discussions and gratefully acknowledge financial support by the
Deutsche Forschungsgemeinschaft through grant SPP 1458.  We thank the
centre for scientific computing (CSC, LOEWE-CSC) in Frankfurt for
computing facilities.

\end{document}